\documentclass[lettersize,journal]{IEEEtran}
\usepackage{url}
\usepackage{amsmath}  % For advanced math typesetting
\usepackage{amssymb}  % For additional math symbols
\usepackage{braket}   % For bra-ket notation
\usepackage{graphicx}
\usepackage{array}

\DeclareMathOperator*{\argmax}{arg\,max}

\begin{document}
\title{Comparison of Feature Learning Methods for Metadata Extraction from PDF Scholarly Documents}

%\iffalse
\author{%
Zeyd Boukhers\textsuperscript{1,2,3}\thanks{\textsuperscript{1}Fraunhofer Institute for Applied Information Technology FIT, Sankt Augustin, Germany.}%
\thanks{\textsuperscript{2}University Hospital of Cologne, Cologne, Germany.}%
\thanks{\textsuperscript{3}University of Koblenz, Koblenz, Germany (e-mail: zeyd.boukhers@fit.fraunhofer.de).}%
\and~
Cong Yang\textsuperscript{4}\thanks{\textsuperscript{4}Soochow University.}%
}

%\fi

%\author{Anonymous Submission}

\markboth{}{Boukhers \MakeLowercase{\textit{et al.}}}

\IEEEtitleabstractindextext{%
\begin{abstract}
The availability of metadata for scientific documents is pivotal in propelling scientific knowledge forward and for adhering to the FAIR principles (i.e. Findability, Accessibility, Interoperability, and Reusability) of research findings. However, the lack of sufficient metadata in published documents, particularly those from smaller and mid-sized publishers, hinders their accessibility. This issue is widespread in some disciplines, such as the German Social Sciences, where publications often employ diverse templates. To address this challenge, our study evaluates various feature learning and prediction methods, including natural language processing (NLP), computer vision (CV), and multimodal approaches, for extracting metadata from documents with high template variance. We aim to improve the accessibility of scientific documents and facilitate their wider use. To support our comparison of these methods, we provide comprehensive experimental results, analyzing their accuracy and efficiency in extracting metadata. Additionally, we provide valuable insights into the strengths and weaknesses of various feature learning and prediction methods, which can guide future research in this field.
\end{abstract}

}

\maketitle

\IEEEdisplaynontitleabstractindextext

\IEEEpeerreviewmaketitle

\section{Introduction}
\label{sec:intro}

%The scientific communities owe much of the flourishing of scientific development nowadays to the easy findability and accessibility of scientific documents. This is due to the availability of scientific metadata which allows those documents to be indexed and linked together in a large and consistent graph. Consequently, Scientometrics as a field of study has been founded to analyse scholarly literature. For most publishers, these metadata are collected directly from the scientific venues and authors alongside the scientific documents to ensure the integrity, completeness and correctness of the data. This means that the document is published in association with its metadata. However, this is unfortunately not the case for some disciplines (e.g. German Social Sciences), whose substantial portion of publications are not easily findable \cite{boukhers2021mexpub, boukhers2022vision}. This is because the main publishers of the publications in these disciplines are small and mid-sized and usually don't have the ability -or are not keen- to provide the metadata, especially for old publications. 

The widespread availability of scientific metadata has greatly contributed to the success and advancement of the scientific community by enabling the easy findability and accessibility of scientific documents. This is achieved by indexing and linking scientific papers in a large and consistent graph such as the OpenAIRE graph~\cite{manghi2022openaire} or the Open Research Knowledge Graph~\cite{jaradeh2019open}. As a result, the field of scientometrics has emerged to study and analyze scholarly literature. While it has become increasingly common for publishers and authors to collect and provide comprehensive metadata alongside the publication of scientific documents, to ensure data's accuracy, completeness, and integrity, this practice was not always popular. Historically, certain disciplines, such as Social Sciences, have seen a considerable portion of their publications become less discoverable due to inadequate metadata collection. This shortfall is particularly evident in works from smaller or mid-sized publishers, which may have lacked the resources or incentive to adequately document metadata, especially in the case of older publications~\cite{boukhers2021mexpub, boukhers2022vision}. Consequently, numerous initiatives have been established to consolidate efforts towards enhancing the findability, accessibility, interoperability, and reusability of scholarly data. Prominent among these are The European Open Science Cloud (EOSC)\footnote{\url{https://eosc-portal.eu/}}, The German Research Data Infrastructure (NFDI)\footnote{\url{https://www.nfdi.de/?lang=en}} and European Strategy Forum on Research Infrastructures (ESFRI)\footnote{\url{https://www.esfri.eu/}}. The primary focus of these initiatives is on the pivotal task of making metadata universally available.

%In most cases, scientific metadata is collected directly from publishers and authors alongside the publication of scientific documents in order to ensure the accuracy, completeness, and integrity of the data. However, this is not always the case for some disciplines, such as German Social Sciences, where a significant portion of published documents may not be easily discoverable due to a lack of metadata, particularly for publications from small or mid-sized publishers who may not have the resources or incentive to provide it, especially for older publications. \cite{boukhers2021mexpub, boukhers2022vision}

%Alternatively, the missing metadata can be made available by extracting them directly from scientific documents. Given the huge number of publications, accomplishing this task manually is laborious and time-consuming, which makes automating the extraction crucial. To this end, several approaches have been proposed to extract metadata from PDF documents using classical Natural Language Processing (NLP)-based approaches~\cite{}. 

Alternatively, the metadata can be directly extracted from scientific documents. However, manually extracting metadata from the vast number of published documents is a labour-intensive and time-consuming task, making automation essential. To automate the process, several approaches have been proposed, including classical natural language processing (NLP)-based approaches~\cite{wang2018clinical, hirschberg2015advances}, which aim to extract metadata from PDF documents efficiently and accurately.

With the recent advances in Deep Neural Networks (DNNs) on textual data, significant results have been achieved on this task~\cite{nayaka2023efficient}. This is due to the capability of these networks to capture latent features from the textual documents. However, the problem is still open and far from being solved because scientific documents come in different templates and layouts. This makes it difficult for any model to find common patterns in the order of the classes. To overcome this problem, some works~\cite{boukhers2021mexpub, ali2023computer} propose to tackle the problem using image processing techniques and taking advantage of the remarkable advances in computer vision. To this end, these techniques view the scientific PDF documents as RGB images. Furthermore, to harness the strengths of both text and visual information, several studies~\cite{boukhers2022vision, balasubramanian2016multimodal} have adopted multimodal approaches, demonstrating notable effectiveness. %Although different approaches have been proposed using different modalities, we found that no approach outperforms the other approaches. 

%In this work, we compare different feature learning and classification approaches for metadata extraction from scientific PDF documents. In addition to comparing the approaches from a technical perspective, we compare their performances and results on a unique and large dataset. The challenge here is that most approaches (i.e. DNN-based) require a very large ground truth dataset for training. Preparing such a dataset is not an easy task as the annotation is laborious and time-consuming. Also, the annotation process is not self-explanatory and often requires quality checks. To overcome this problem, we synthetically generated *** samples using a manually predefined set of templates and already available metadata. In summary, the contributions of this paper are as follows: 
This study explores a variety of feature learning and classification approaches to extract metadata from scientific PDF documents, emphasizing the use of methodologies best suited to the specific challenges of this task. We employ classical approaches such as Conditional Random Fields, advanced NLP techniques including BiLSTM with BERT representations, and innovative multimodal and Textmap methods. While generative LLMs like GPT-4 or LLAMA excel in natural language generation, they are not ideal for structured tasks such as metadata extraction from scientific PDFs. These models, designed primarily for text generation from prompts, face difficulties with fixed formats, which can lead to inaccuracies from over-generalization and context sensitivity, and require substantial resources for task-specific tuning. By contrast, our chosen approaches leverage the strengths of BERT and other architectures to efficiently handle the unique layout variability and multimodal content of scientific documents, ensuring precise and reliable metadata extraction.

In addition to evaluating the technical aspects of these approaches, we also compare their performance and results on a large and unique dataset. One challenge in this area is that many techniques, such as those based on deep neural networks (DNNs), require an extensive ground truth dataset for training. However, creating such a dataset can be difficult, as the process of annotating the data is time-consuming and labor-intensive, and often requires quality checks. To address this issue, we created two challenging datasets, namely, \emph{SSOAR-MVD} and \emph{S-PMRD}. For \emph{SSOAR-MVD}, we synthesized 50.000 samples using a predefined set of templates and available metadata. \emph{S-PMRD} is an authentic subset of the Semantic Scholar Open Research Corpus. The main contributions of this paper are as follows:

\begin{itemize}

    \item We present a variety of approaches for extracting metadata from scientific PDF documents.
    \item We created a large, labelled dataset for metadata extraction from scientific PDF documents.
    \item We conducted extensive experiments to compare the various approaches.
    \item To facilitate reproducibility and future development, we have made the implementations of all the approaches publicly available\footnote{\url{Will be released upon publicaiton.}}.
\end{itemize}

The remainder of this paper is organized as follows: In Section~\ref{sec:rew}, we review related works. In Section~\ref{sec:app}, we introduce all the approaches covered in this paper. Section~\ref{sec:exp} presents the dataset and experimental results, and finally, in Section~\ref{sec:con}, we provide concluding remarks and discuss potential future directions.
\section{Related Work}
\label{sec:rew}
Metadata extraction, while a specialized subset of information extraction (IE), serves a distinct purpose and presents unique challenges. This section provides an overview of the most pertinent techniques for metadata extraction, categorizing them into three distinct groups for a clearer understanding of their applications and methodologies.

\subsection{Natural Language Processing}

Metadata extraction in Natural Language Processing (NLP) has primarily been approached through two distinct methodologies: rule-based and machine learning-based techniques \cite{DAY2007152}. Rule-based techniques rely on predefined rules developed through human expertise to guide metadata extraction \cite{DAY2007152}. These methods are generally more straightforward to implement but may lack the adaptability found in machine learning-based systems~\cite{Yingsaeree2005, 10.1145/1066677.1066917}. On the other hand, machine learning-based approaches, exemplified by platforms like CiteSeerX \cite{li2006citeseerx}, leverage supervised learning algorithms trained on labelled datasets to autonomously extract metadata from new documents. These algorithms range from Hidden Markov Models (HMM) \cite{Kristie:1999}, Conditional Random Fields (CRFs) \cite{Peng2004AccurateIE}, to Support Vector Machines (SVM) \cite{Hui:2003}. Although robust and effective, the drawback of these machine learning methods lies in the labour-intensive labelling of training data, especially when dealing with samples of high variability.

Recent advancements in Deep Neural Networks (DNN) have provided a new dimension to the field of metadata extraction. DNNs have been shown to considerably outperform traditional methods in effectiveness and efficiency \cite{DAY2007152}. \cite{Zhiheng:2015} pioneered a Bidirectional LSTM-CRF model, combining Long Short-Term Memory (LSTM) with a Conditional Random Field (CRF) layer to encode word sequences and predict labels. Similarly, \cite{Jason:2016} employed a Bidirectional LSTM integrated with a Convolutional Neural Network (CNN) to generate character-level word representations. \cite{Dong:2017} introduced a DNN-based Segment Sequence Labeling for metadata extraction, setting new performance benchmarks. This approach outstripped existing works such as ParsCit \cite{ParsCit}, a CRF-based model, and BibPro \cite{Bibpro}, a neural network-based model, when evaluated on public datasets like UMass \cite{UMass} and Cora \cite{Kristie:1999}.

\subsection{Computer Vision}

While Computer Vision (CV) approaches are not yet ubiquitously applied in the field of metadata extraction, emerging research indicates their promising capabilities, especially for Natural Language Processing (NLP) related tasks. One notable example is DeepPDF \cite{Stahl:2018}, which applies a unique perspective to PDF document segmentation. Instead of traditional text-based analysis, DeepPDF treats the document as an image and employs UNet-Zoo, a specialized architecture originally designed for biomedical image segmentation. This approach allows for accurate paragraph identification while ignoring other elements like headers, captions, figures, and references, thus substantiating the potential of CV-based techniques for textual document analysis.

Building upon the groundwork laid by \cite{Stahl:2018}, MexPub \cite{Mexpub} introduced an innovative technique for extracting metadata from German PDF documents. The methodology utilizes a pixel-by-pixel analysis through the MASK-RCNN architecture \cite{8237584}, specifically engineered for object detection and classification. It incorporates the ResNeXt backbone \cite{Xie2016} and Feature Pyramid Networks (FPN) for feature extraction from raw images. While MexPub has shown promising results, it encounters limitations in certain areas. For example, the model struggles with generalizing to scientific literature that diverges structurally from the training dataset. Additionally, MexPub faces challenges in precisely detecting smaller patterns or those placed in unconventional positions. These limitations suggest that the method's performance could be further enhanced by incorporating text processing elements into a unified architecture.

\subsection{Multimodality}

Multimodal deep learning has made significant inroads across various applications, including but not limited to audiovisual and image classification, showcasing impressive performance. Specifically within the realm of metadata extraction, there's growing evidence that multimodal approaches are superior to their unimodal counterparts, as highlighted in studies by \cite{Balasubramanian:2015}, \cite{Liu:2018}, and \cite{boukhers2022vision}.

Balasubramanian et al. \cite{Balasubramanian:2015} employed a combined audio and video modality strategy to extract metadata from video lectures. Their technique harnessed the potential of a Naive Bayes classifier in tandem with a rule-based refiner. The essence of this methodology was capitalizing on the interplay between audio transcripts and the content of slides embedded within video streams. Astonishingly, this synergy yielded a marked $114.2\%$ improvement in metrics such as F-score, precision, and recall when benchmarked against solely audio-based methodologies.

Liu et al. \cite{Liu:2018} pioneered a multimodal deep-learning strategy tailored for metadata extraction from scientific documents. Their model seamlessly ingests both image and textual data, negating the need for handcrafted classification features. On the textual front, Recurrent Neural Networks (RNNs) were employed, while image data was processed using Convolutional Neural Networks (CNNs). The amalgamated representation was then processed via a BiLSTM network, culminating in classification through a CRF classifier. The potency of this composite approach was evident when juxtaposed against unimodal strategies.

Further enriching the field,\cite{boukhers2022vision} presented an intriguing approach to address metadata extraction challenges specific to German scientific papers, which frequently exhibit a vast array of layouts due to the varied publishing standards of small to mid-sized publishers. The paper proposed a multimodal approach that perceives a PDF document simultaneously as an RGB image and a textual document, using BiLSTM and MexPub, respectively. The outputs from both sub-models are subsequently merged and processed by another BiLSTM model for token classification.

\section{Approach}
\label{sec:app}
%In this section, we describe the different feature learning and classification methods for extracting metadata from scientific PDF documents. Similarly to most studies tackling this problem, we assume that only the first page of the PDF document can contain these metadata and that their availability varies across documents. For instance, \emph{DOI} is not supposed to occur in all scientific PDF documents.

This section discusses various feature learning and classification methods for extracting metadata from scientific PDF documents. Like many studies in this area, we assume that metadata may only be present on the first page of a PDF document and that its availability may vary across documents. For example, all scientific PDF documents may not include the Digital Object Identifier (DOI).

Let $\mathcal{P}$ be the first page of a scientific PDF document, consisting of a set of observed words $\mathbf{\omega}=\langle\omega_1,\omega_2,\cdots,\omega_{n}\rangle$, where $n=\lvert\mathbf{\omega}\rvert$. Let $S$ be a set of states in a finite state machine, each corresponding to a label $l\in L$ (e.g., \emph{Title}, \emph{Authors}, etc.). The task is to formalize $\gamma (\mathcal{P})=\mathbf{s}$, where $\mathbf{s}=\langle s_1,s_2,\cdots,s_n\rangle$ is the sequence of states in $S$ that correspond to the labels assigned to the words in the input sequence $\mathbf{\omega}$. Table~\ref{tab:variables} represents the used variables and their descriptions

\begin{table}[ht!]
    \centering
    \begin{tabular}{|c|p{0.7\linewidth}|}
    \hline
      \textbf{Variable}   & \textbf{Description} \\
      \hline
       $\gamma$  & The metadata extraction model\\
       \hline
       $\mathcal{P}$ & The first page of the PDF document \\
       \hline 
       $\mathbf{S}$ & The outcome of the model, which is a set of strings associated with their labels\\
       \hline
       $y$ & The metadata label\\
       \hline   
       $\mathbf{s}_i$ & The output metadata value of the $y$th label\\
       \hline
       $K$ & Section \\
       \hline 
       $w$ & Classified token\\
       \hline
       $\omega$ & Unclassified token \\
       \hline
    \end{tabular}
    \caption{Overview of key variables used in this paper across the different feature learning and classification methods for metadata extraction.}
    \label{tab:variables}
\end{table}

%In this study, we compare a range of techniques for extracting metadata from scientific PDF documents. In addition to introducing new methods (described in Sections~\ref{sec:rcnn}, \ref{sec:multimodal} and \ref{sec:tmp}), we also briefly summarize existing approaches (detailed in Sections~\ref{sec:crf}, \ref{sec:bilstm}, \ref{sec:bilstm-crf} and \ref{sec:grobid}). All techniques are evaluated following the aforementioned formalization of the metadata extraction task.

In this study, we compare several approaches for extracting metadata from scientific PDF documents, including foundational techniques like Conditional Random Fields (CRF)~\cite{souza2014arctic} and GROBID~\cite{GROBID}, which have established the groundwork for metadata extraction. We also implement and explore novel neural sequence labeling approaches using BiLSTM and BiLSTM-CRF architectures (Sections~\ref{sec:bilstm} and \ref{sec:bilstm-crf}). This work introduces three new methodologies that take different approaches to the problem: a computer vision approach using Fast R-CNN (Section~\ref{sec:rcnn}), a multimodal neural architecture (Section~\ref{sec:multimodal}), and our proposed TextMap framework (Section~\ref{sec:tmp}). All approaches are evaluated following the aforementioned formalization of the metadata extraction task, enabling a comprehensive comparison of their effectiveness.

\subsection{Conditional Random Fields (CRF)\cite{souza2014arctic}}
\label{sec:crf}

%This approach proposed by XYZ employs a two-layer CRF model aiming to reduce the complexity of the task. Given the extracted lines from $\mathcal{P}$, the first layer classifies each one into one of the five main sections with metadata information (Header, Title, Author Information, Body and Footnote). To this end, font features such as size and alignment are extracted from each line and fed to the CRF layer. From each of the identified sections, the respective metadata is extracted. This means that some metadata is already excluded from some sections. Given a section $j$, each word $w_{i,j}$ is represented with handcrafted features (e.g. length, following year format, etc.). Subsequently, the sequential feature vectors are fed into the second CRF layer to recognize the metadata. 

\begin{figure}
    \centering
    \includegraphics[width=\linewidth]{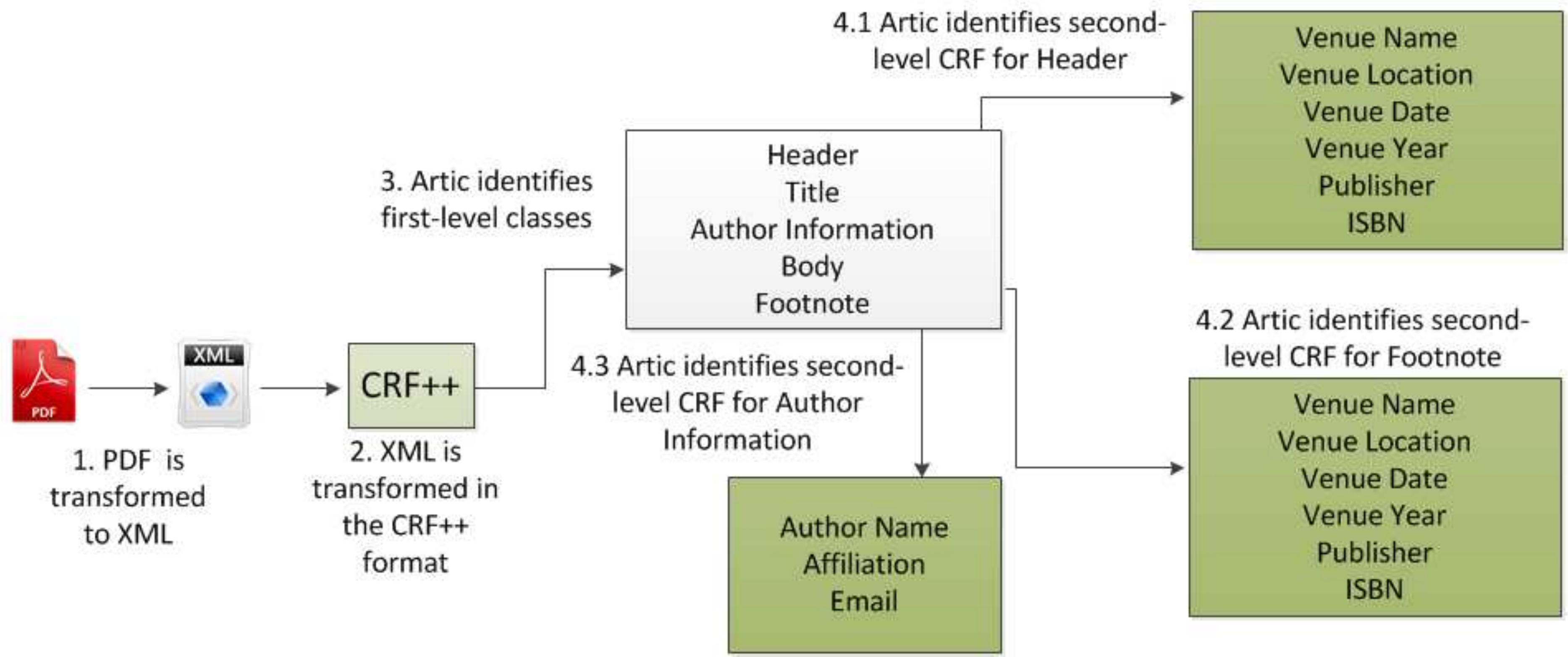}
    \caption{Schematic representation of the two-layer Conditional Random Field (CRF) model for metadata extraction~\cite{souza2014arctic}.}
    \label{fig:crf}
\end{figure}

The approach proposed by Souza, Viviane, and Heuser~\cite{souza2014arctic} employs a two-layer Conditional Random Field (CRF) model for extracting metadata from scientific PDF documents. As illustrated in Figure~\ref{fig:crf}, the extraction process is divided into two main steps: identifying main sections and extracting metadata from these sections.

Given the extracted lines from the first page $\mathcal{P}$, the first layer of the CRF model classifies each line into one of the five main sections that may contain metadata information: \emph{Header}, \emph{Title}, \emph{Author Information}, \emph{Body}, and \emph{Footnote}. To achieve this, the model processes font features such as size, style, and alignment from each line and uses them as input. Once the main sections have been identified, the second layer of the CRF model is responsible for extracting metadata from these sections. Some content is automatically excluded from certain sections during this process. For instance, content that appears in the Footnote section would not be included in the model's output.

For each identified section $k$, the model processes a sequence of observed words $\mathbf{\omega}=\langle\omega_1,\omega_2,\cdots,\omega_{n}\rangle$. Each word $\omega_i$ is represented with a feature vector $\mathbf{x}_i$, which comprises $m$ handcrafted features such as length, whether it follows a year format, presence of special characters, and capitalization, among others. The model calculates the probability of a section sequence given a handcrafted feature sequence using the following equation:

\begin{equation}
P(\mathbf{s} \mid \mathbf{x}) = \frac{\exp\left(\sum_{i=1}^n\sum_{j=1}^m\lambda_j f_j(y_{i-1},y_i,\textbf{x},i)\right)}{Z(\mathbf{x})}
\label{eq:crf1}
\end{equation}

\noindent where $f_j$ denotes the feature function for the $j^{th}$ feature, and $\lambda_j$ is the corresponding weight parameter. $Z(\mathbf{x})$ is the normalization factor, ensuring that the sum of probabilities over all possible label sequences equals $1$:

\begin{equation}
Z(\mathbf{x}) = \sum_{y'} \exp\left(\sum_{i=1}^n\sum_{j=1}^m\lambda_j f_j(y'_{i-1},y'_i,\textbf{x},i)\right)
\end{equation}

To find the optimal weights ${\lambda}_{j=1}^m$, a training process is conducted by maximizing the log-likelihood of the training data:

\begin{equation}
\mathcal{L}(\lambda)=\sum_{u=1}^{\lvert D \lvert} \log P\left(y^{(u)},\mathbf{x}^{(u)}\right)-\frac{\sum_{j=1}^m \lambda_j^2}{2\sigma^2}
\end{equation}

\noindent where $(x^{(u)}, y^{(u)})$ are the pair features and label of the $u^{th}$ training instance in the training dataset $D$, and $\sigma^2$ is a hyperparameter for L2 regularization  that controls the model’s complexity.

\subsection{Bi-Directional LSTM}
\label{sec:bilstm}

For this solution, we employed a Bidirectional Long Short-Term Memory (BiLSTM) model with three layers. The BiLSTM has 112 hidden dimensions and is followed by two fully connected layers. The final layer uses a softmax activation function to assign each word to a specific class. Given a sequence of observed words $\mathbf{\omega}=\langle\omega_1,\omega_2,\cdots,\omega_{n}\rangle$, the embedding vector of each word $\omega_i$ is obtained the BERT model:

\begin{figure}
    \centering
    \includegraphics[width=0.85\linewidth]{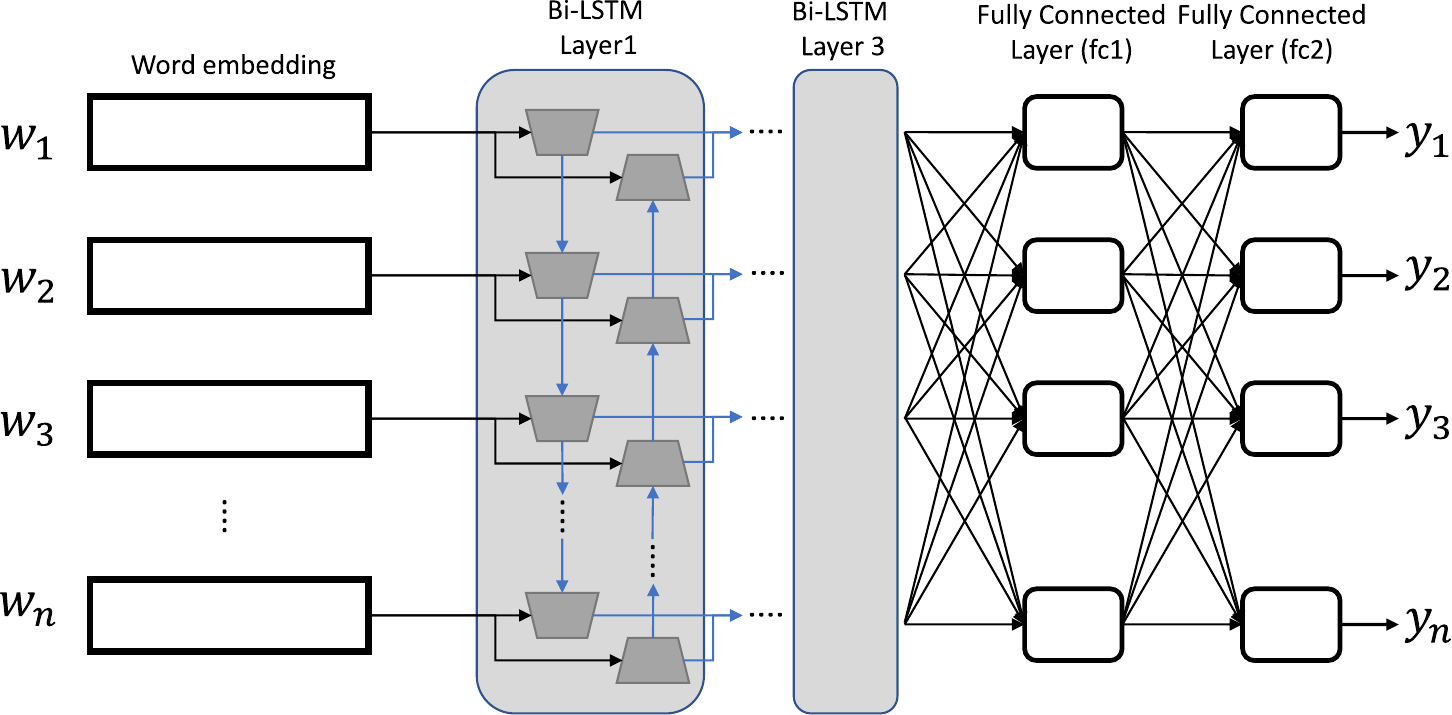}
    \caption{Diagram of the Bi-Directional LSTM network architecture for metadata extraction}
    \label{fig:bi-lastm}
\end{figure}

\begin{equation}
    \mathbf{x_i}= \textrm{BERT}(\omega_i), \quad i=1,2,\cdots, n
    \label{eq:bilstm1}
\end{equation}

\noindent resulting in a sequence of embedding vectors $\mathbf{x}=\langle\mathbf{x}_1,\mathbf{x}_2,\cdots,\mathbf{x}_{n}\rangle$.

This sequence of embedding vectors is fed into the three bidirectional LSTM layers with hidden dimensions $=112$.  Let $\mathbf{h}^{(f)}_i$ and $\mathbf{h}^{(b)}_i$ denote the forward and backward hidden states at position $i$ in the sequence. The hidden states are updated as follows:

\begin{equation}
\begin{split}
    \mathbf{h}^{(f)}_i=\textrm{LSTM}^{(f)} (\mathbf{x}_i, \mathbf{h}^{(f)}_{i-1}) ,\\
    \mathbf{h}^{(b)}_i=\textrm{LSTM}^{(b)} (\mathbf{x}_i, \mathbf{h}^{(b)}_{i+1})
\end{split}   
\label{eq:bilstm2}
\end{equation}

For each BiLSTM layer $t$, the outputs of the forward and backward LSTM units are concatenated to form the hidden state of the BiLSTM layer:

\begin{equation}
    \mathbf{h}^{(t)}_i=\left[\mathbf{h}^{(f,t)}_i; \mathbf{h}^{(b,t)}_i  \right]
    \label{eq:bilstm3}
\end{equation}

The output of the last BiLSTM layer is passed through two fully connected layers with weight matrices $\mathbf{W}_1$ and $\mathbf{W}_2$ and bias vectors $\mathbf{b}_1$ and $\mathbf{b}_2$:

\begin{equation}
    \mathbf{o}_i=\textrm{ReLU}(\mathbf{W}_2 \textrm{ReLU} (\mathbf{W}_1 \mathbf{h}^{(\textrm{BiLSTM})}_i + \mathbf{b_1}) + \mathbf{b_2}) 
\end{equation}

A softmax activation function is applied to the output of the last fully connected layer to compute the probability distribution over the predefined set of labels for each word in the sequence:

\begin{equation}
    \hat{\mathbf{y}}_i=\textrm{SoftMax}(\mathbf{o}_i) = \frac{\exp (\mathbf{o}_i) }{\sum_{l=1}^L \exp (\mathbf{o}_i, l)}
\end{equation}
\subsection{BiLSTM-CRF}
\label{sec:bilstm-crf}

\begin{figure}
    \centering
    \includegraphics[width=0.85\linewidth]{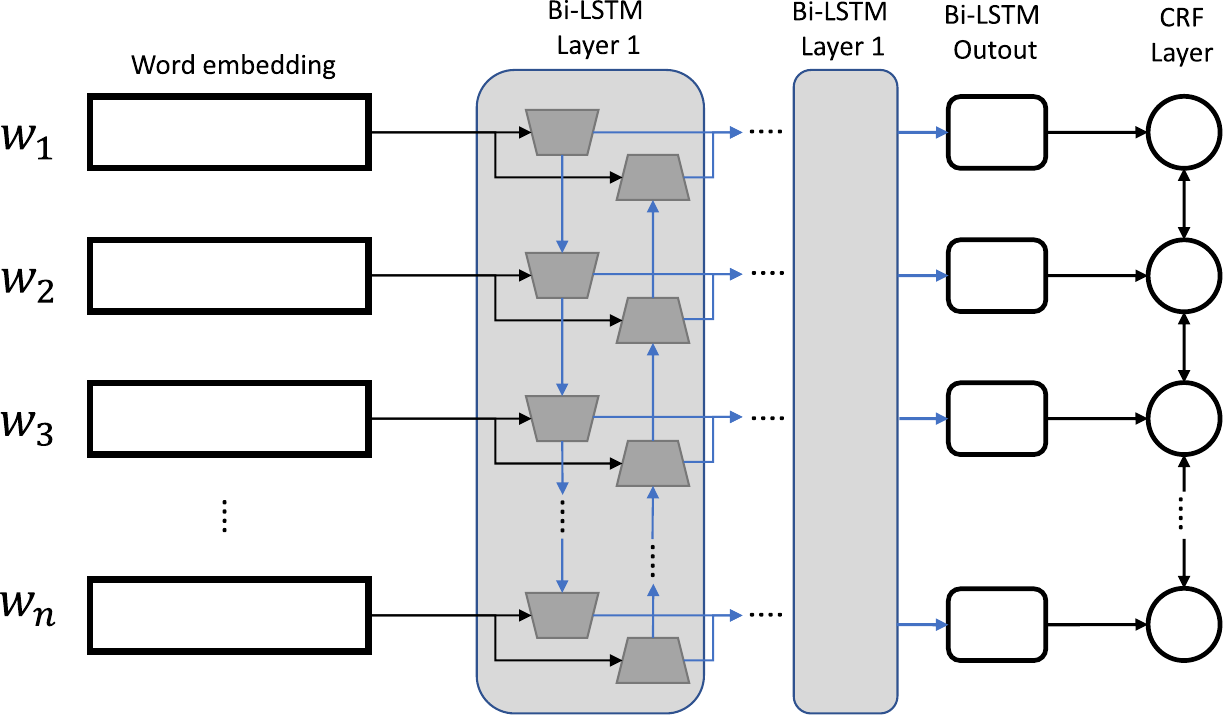}
    \caption{Diagram of the Bi-Directional LSTM network architecture with CRF classifier for metadata extraction}
    \label{fig:bi-lastm_crf}
\end{figure}

As BiLSTM-CRF is used in many NLP tasks and specifically extracting information from textual data~\cite{alzaidy2019bi,dai2019grantextractor}, we assume that it would perform similarly on the task of extracting metadata from PDF documents. Figure~\ref{fig:bi-lastm_crf} illustrates the developed model that takes as input the embeddings of the words extracted from $\mathcal{P}$. The embeddings are obtained using a pre-trained BERT model. The assumption is that most of the metadata classes are represented in structured phrases that BERT can capture. For the other classes (e.g. Author name), 

The proposed model consists of a 4-layer BiLSTM network with 115 hidden dimensions followed by a CRF layer for sequence labelling. The sequence of observed words goes through the same steps as mentioned in the Equations~\ref{eq:bilstm1}, \ref{eq:bilstm2} and \ref{eq:bilstm3}. Then, the output of the last BiLSTM layer $H=\sum_i \mathbf{h}^{\textrm{BiLSTM}}_i$ is passed through a CRF layer to calculate the probability of a label sequence $P(\mathbf{s} \mid H)$,  using Equation~\ref{eq:crf1}.

\subsection{Grobid\cite{GROBID}}
\label{sec:grobid}

GROBID is a machine-learning library that is designed to extract, parse, or restructure raw documents into structured XML/TEI documents. It employs a cascade of sequence labelling models to parse each document, allowing it to adapt to the different hierarchical structures present in the documents. By utilizing a cascade approach, GROBID can handle a wide variety of document layouts and structures.

The main idea behind GROBID's approach is to break down the complex task of document parsing into a series of smaller, more manageable tasks. Each model in the cascade focuses on a specific aspect of the document structure, such as headers, titles, author information, or other metadata. The models have a small number of labels, which makes it easier to manage and train. When combined, the full cascade provides a detailed end-result structure.

In GROBID, the models are organized hierarchically to address the inherent hierarchical structure of the documents. The original GROBID model produces 55 different "leaf" labels, which are the final labels assigned to the text elements after the document has been processed by the entire cascade of models. Each "leaf" label corresponds to a specific element in the structured XML/TEI output.

\begin{figure}
    \centering
    \includegraphics[width=\linewidth]{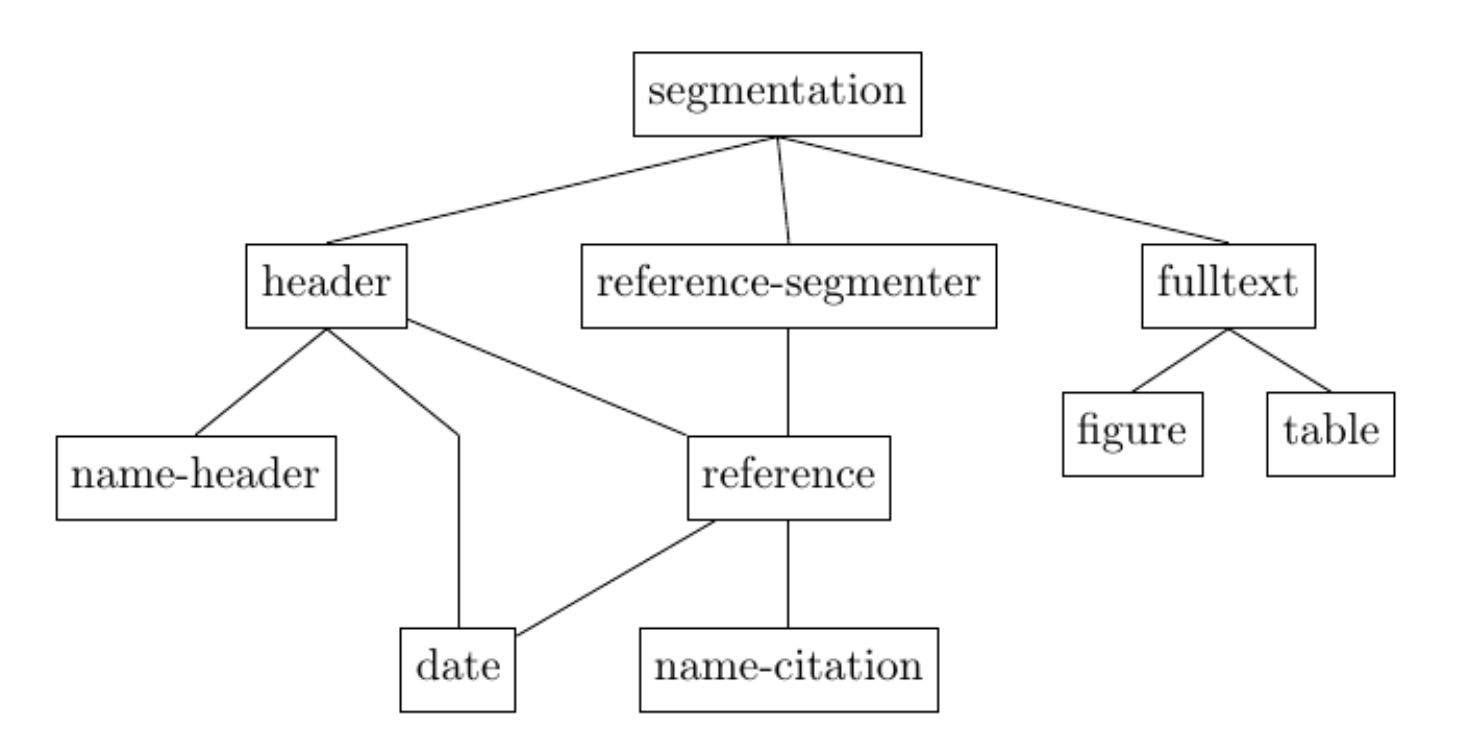}
    \caption{Overview of the GROBID Framework for structured metadata extraction \cite{GROBID}}
    \label{fig:my_label}
\end{figure}

\textbf{Model Training}: Train a cascade of sequence labeling models on the training dataset. Each model in the cascade is responsible for recognizing and classifying specific elements of the document structure. The models are organized hierarchically, with each model feeding its output to the subsequent model in the cascade.

\textbf{Model Inference}: Given a new document, apply the trained cascade of models to parse the document and extract metadata. The output of each model is fed into the next model in the cascade, refining the document structure at each step. Finally, the "leaf" labels are assigned to the text elements, resulting in a structured XML/TEI representation of the document.

\textbf{Metadata Extraction}: Once the document has been parsed and structured, the metadata can be easily extracted from the XML/TEI representation by querying the relevant elements and their associated "leaf" labels.
\subsection{Fast-RCNN}
\label{sec:rcnn}

In earlier work~\cite{boukhers2021mexpub}, we addressed this problem by viewing the PDF document as an image and leverage from the advanced progress in computer vision. 

The model is an adaptation of Mask R-CNN, a cutting-edge object instance segmentation technique proposed by He et al~\cite{he2017mask}. It identifies objects within images at the pixel level by extending Faster RCNN with an additional branch for predicting object masks and utilizing Region of Interest (RoI)-Align instead of RoI-Pooling. The binary object mask highlights the position of each object in its bounding box on a pixel-by-pixel basis. In this implementation, Mask R-CNN is combined with a ResNeXt~\cite{xie2017aggregated} backbone architecture and a Feature Pyramid Network (FPN), following the approach, outlined in~\cite{lin2017feature}.

\begin{figure*}[ht!]
    \centering
    \includegraphics[width=\linewidth]{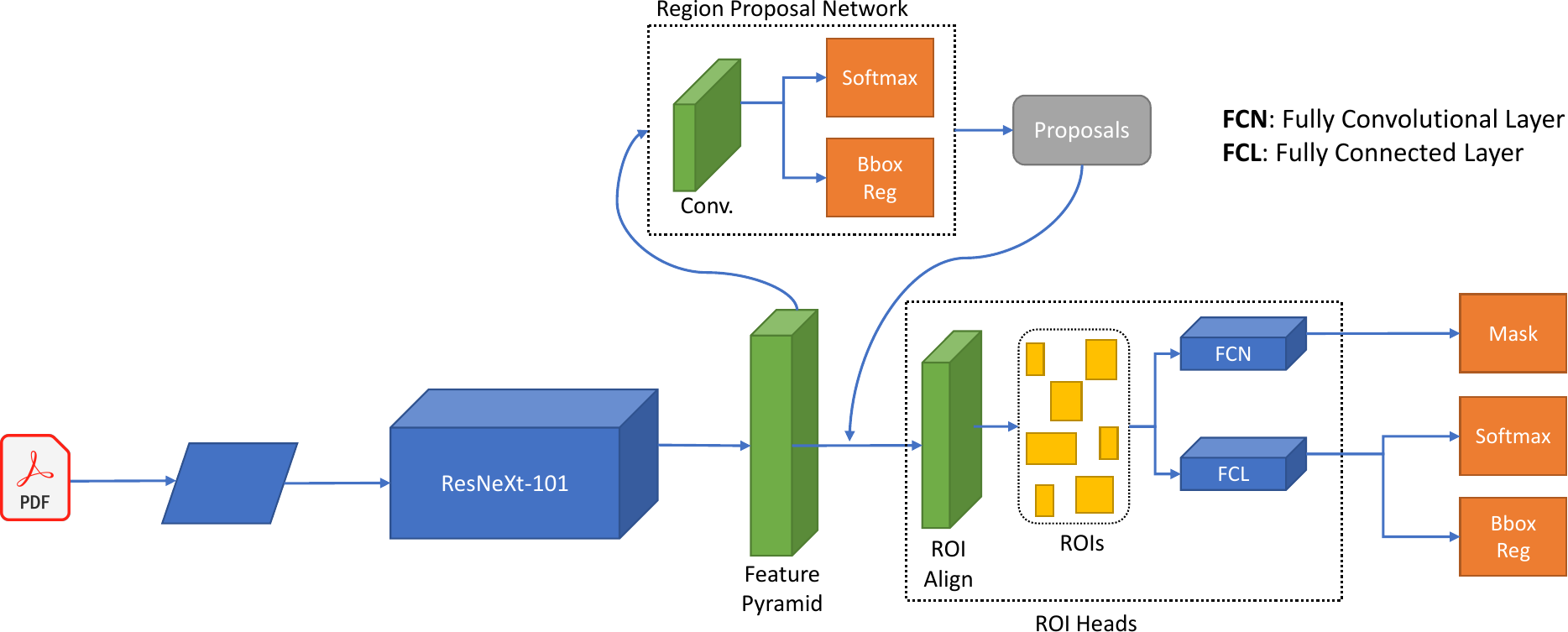}
    \caption{Mask R-CNN architecture employed for metadata extraction from PDF pages.}
    \label{fig:maskrcnn}
\end{figure*}

As illustrated in Figure~\ref{fig:maskrcnn}, the PDF page $\mathcal{P}$ is first transformed into a pixel image, which serves as input for the RCNN model. The model is composed of three main components: (i) a Feature Pyramid Network (FPN) with ResNeXt as a backbone network, (ii) a Region Proposal Network (RPN), and (iii) RoI (Region of Interest) Heads. As detailed in Table\ref{tab:ResNeXt}, the ResNeXt backbone includes a stem block and four stages, each containing multiple bottleneck blocks.

The stem block down-samples the input image twice through a $7\times7$ convolution with a stride of 2, and max-pooling with a stride of 2, generating a feature map at a 1/4 scale. The subsequent four stages contain bottleneck blocks, each featuring three convolutional layers with kernel sizes of $1\times1$, $3\times3$, and $1\times$1. These stages consist of 3, 4, 23, and 3 bottleneck blocks respectively, and produce feature maps at scales of $1/4$, $1/8$, $1/16$, and $1/32$~\cite{xie2017aggregated}. A max-pooling layer with a kernel size of 1 and a stride of 2 is introduced to the final stage of ResNeXt, yielding a feature map at a $1/64$ scale~\cite{he2017mask}.

The second component, the Region Proposal Network (RPN), suggests candidate object bounding boxes utilizing the outputs from the FPN's five stages. Subsequently, a fully convolutional mask prediction branch is integrated into the head~\cite{he2017mask}. The RoI head employs fully-connected layers to generate refined box locations and classification results from multiple fixed-size features, which are obtained by cropping and warping feature maps. The box head then filters out up to 100 boxes using non-maximum suppression (NMS) to eliminate redundant detections.

\begin{table}[]
    \centering
    \begin{tabular}{c|c|c|c}
        Layer name & scale &  kernel size &  stride\\
        \hline
        \hline
        stem & $1/4$ & $7\times7$ & 2\\
        \hline
        backbone 1 & $1/4$ & $\begin{bmatrix} 1 & \times & 1\\ 3 & \times & 3 \\ 1 & \times & 1\end{bmatrix}\times 3$ & 1\\
        \hline
        backbone 2 & $1/8$ & $\begin{bmatrix} 1 & \times & 1\\ 3 & \times & 3 \\ 1 & \times & 1\end{bmatrix}\times 4$ & 1\\
        \hline
        backbone 3 & $1/16$ & $\begin{bmatrix} 1 & \times & 1\\ 3 & \times & 3 \\ 1 & \times & 1\end{bmatrix}\times 23$ & 1\\
        \hline
        backbone 4 & $1/32$ & $\begin{bmatrix} 1 & \times & 1\\ 3 & \times & 3 \\ 1 & \times & 1\end{bmatrix}\times 3$ & 1\\
        \hline
        \hline
        max pooling layer & $1/64$ & $1 \times 1$ & 2\\
        \hline
    \end{tabular}
    \caption{Overview of the ResNeXt structure in the RCNN model, highlighting the number of bottleneck blocks and the scales of feature maps at each stage.}
    \label{tab:ResNeXt}
\end{table}

Transfer learning is a widely used technique in deep learning for computer vision tasks. It involves retraining pre-trained convolutional networks on smaller, task-specific datasets to fine-tune the weights and biases, leveraging the knowledge gained from one classification task to another~\cite{pan2010survey}. In our study, we employ a source model based on the Detectron2~\cite{wu2019detectron2} implementation of Mask R-CNN ResNeXt-101 32x8d FPN. This model was initially fine-tuned on 191,832 images from the PubLayNet dataset~\cite{zhong2019publaynet}, which includes annotated images of articles from PubMed Central™ Open Access (PMCOA) featuring five classes: title, text, list, table, and figure. The model is well-suited for extracting metadata from scientific papers since it (i) has a backbone trained on the extensive COCO dataset, (ii) underwent fine-tuning on a large dataset of scientific document images, and (iii) is designed for a task closely related to ours.

To adapt this model for extracting metadata patterns from scientific documents, we first modified the final layer of the source model to output nine target classes (title, authors, journal, abstract, date, DOI, address, affiliation, and email addresses) instead of the original five. Empirical experiments on a subset of 103 random samples from our training dataset showed that the best-performing architecture has two frozen layers and 15k iterations. Based on these findings, we fine-tuned the model using the full training dataset, setting the learning rate to $2.5 \times 10^{(-3)}$.

\subsection{Vision and Natural Language}
\label{sec:multimodal}

In earlier work~\cite{boukhers2022vision}, we addressed this problem using a multimodal neural network model that employs NLP together with Computer Vision for metadata extraction. 

\begin{figure*}
    \centering
    \includegraphics[width=\linewidth]{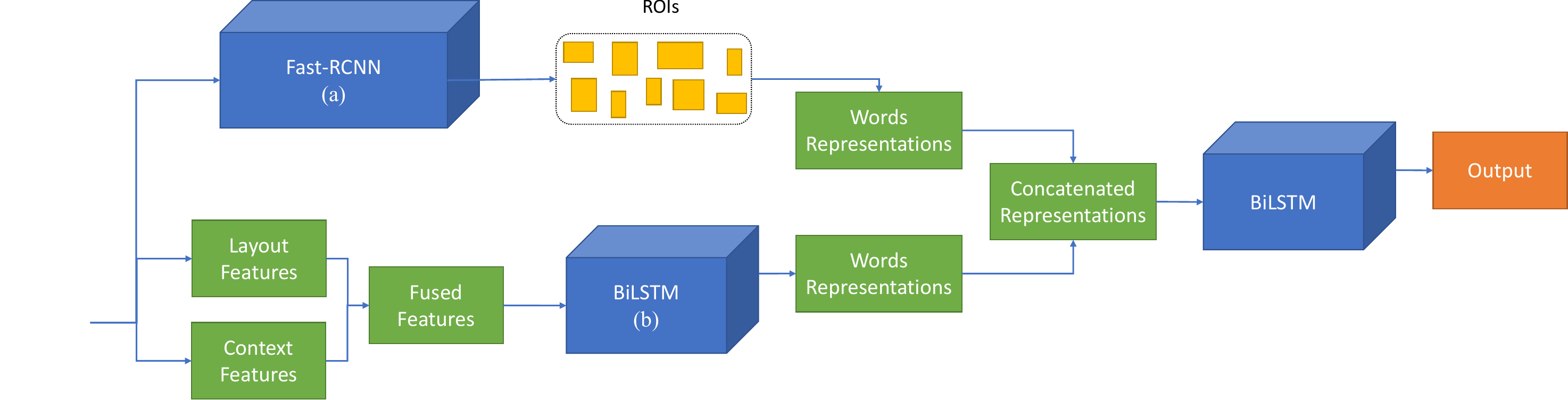}
    \caption{Multimodal extraction approach, where \emph{(a)} refers to the model described in section~\ref{sec:rcnn} and \emph{(b)} refers to the model described in section~\ref{sec:bilstm}}
    \label{fig:multi}
\end{figure*}

Figure~\ref{fig:multi} illustrates the initial step of our process, wherein the text is extracted from $\mathcal{P}$ using CERMINE~\cite{CERMINE}. Known for its reliability in handling diverse layouts at the line level, CERMINE also provides geometric structural information such as text position and font style.

From each extracted token $\omega$, a set of 16 handcrafted features, denoted as $F_{hand}$, is derived. A word embedding for the token, denoted as $F_{embed}$, is also generated, which encapsulates the context and meaning of the words. These two sets of features are then concatenated to form a single feature vector, such that $F_{total} = Concat(F_{hand}, F_{embed})$.

The consolidated vector $F_{total}$ is used as the input for the Natural Language Processing (NLP) sub-model, described in section~\ref{sec:bilstm}. Simultaneously, the image of $\mathcal{P}$ is supplied as input to the Computer Vision (CV) model, described in section~\ref{sec:rcnn}.

\subsubsection{Natural Language-based Model}

To model the extracted text, we utilized a BiDirectional Long-Short-Term Memory (BiLSTM) due to its proven accuracy in handling textual data, as detailed in section~\ref{sec:bilstm}. This sub-model comprises two layers of LSTM models, each with 256 hidden dimensions; the first layer is a forward LSTM, and the second layer is a backward LSTM. Please refer to section~\ref{sec:bilstm} for more details about BiLSTM

The input to this model is a word representation vector with a length of 1041. As previously described, this vector is the concatenation of two vectors. The first vector, consisting of 16 units, encapsulates layout features such as the font size of the word, font style, the spacing between the word and the line above or below it, and flags denoting whether the text is italicized, bolded, or adheres to a specific common format like date or email, among others. The second vector contains the ELMO~\cite{che2018towards} embedding results, derived from a model trained on German documents.

Following the two LSTM layers, a fully connected layer of 512 units is in place, ending with an output layer of 10 neurons, representing the metadata classes. The output layer employs a softmax activation function to generate probability scores for the word's affiliation with each of the classes.

\subsubsection{Computer Vision-based Model}

Building on the proven efficiency of MexPub~\cite{boukhers2021mexpub}, detailed in section~\ref{sec:rcnn}, we leverage it as the Computer Vision (CV) sub-model, feeding it with $\mathcal{P}$. This model yields output in the form of bounding boxes labelled with metadata classes. Subsequently, we extract the text enclosed within the bounding boxes as identified by the CV sub-model and compile the probabilities for all potential classes within that box prior to their submission to the classifier. It's important to note that the CV sub-model may also generate bounding boxes that are unclassified, meaning they do not associate with any of the predefined classes.

\subsubsection{Classifier}
In the final stage of our architecture pipeline, the output of the NLP and CV sub-models is fused using a SoftMax classifier. Specifically, all words from the document are extracted and sequentially traversed. Their vector representations, generated by the sub-models, are then concatenated. A BiLSTM, notable for its bidirectional operation and capability to preserve information from both past and future states, is employed in this context as well. This is especially advantageous for understanding context and discerning patterns within sentences or paragraphs (e.g., if the adjacent words are titles, the current word is highly likely to be a title as well). The model specifically takes in a vector of length 20, resulting from the concatenation of both sub-model outputs. The model's output is a probability distribution of length 10 corresponding to all classes. As depicted in Figure 2, the classifier comprises two stacked LSTM layers (forward and backward LSTMs) each with 256 hidden dimensions. A fully connected layer follows these two layers, encompassing 512 input nodes and 10 output nodes activated by a SoftMax function.

%%%%%%%%%%%%%%%%%%%%%%%%%%%%%%%%%%%%

\subsection{Text Map Approach}\label{sec:tmp}
The text map approach presents a novel framework that jointly optimizes spatial and semantic information for metadata extraction. Given the first page $\mathcal{P}$ of a PDF document and its sequence of observed words $\mathbf{\omega}=\langle\omega_1,\omega_2,\cdots,\omega_{n}\rangle$, our goal is to learn a mapping function $\gamma$ that assigns metadata labels while preserving both spatial and semantic relationships.

%We formulate the metadata extraction problem as an optimization task that maximizes the mutual information between the input document and the extracted metadata labels. The information content $I$ of our extraction function $\gamma$ is bounded by:

%\begin{equation}
    %I(\gamma(\mathcal{P})) \leq I(\mathcal{T}(\mathbf{\omega})) + I(\mathcal{L}(\mathcal{P})) - I(\mathcal{T}(\mathbf{\omega}) \cap \mathcal{L}(\mathcal{P}))
%\end{equation}

\subsubsection{Two Phase Processing}
The approach processes documents through two complementary phases as depicted in Figure~\ref{fig:textmap}:

\begin{figure*}
    \centering
    \includegraphics[width=\linewidth]{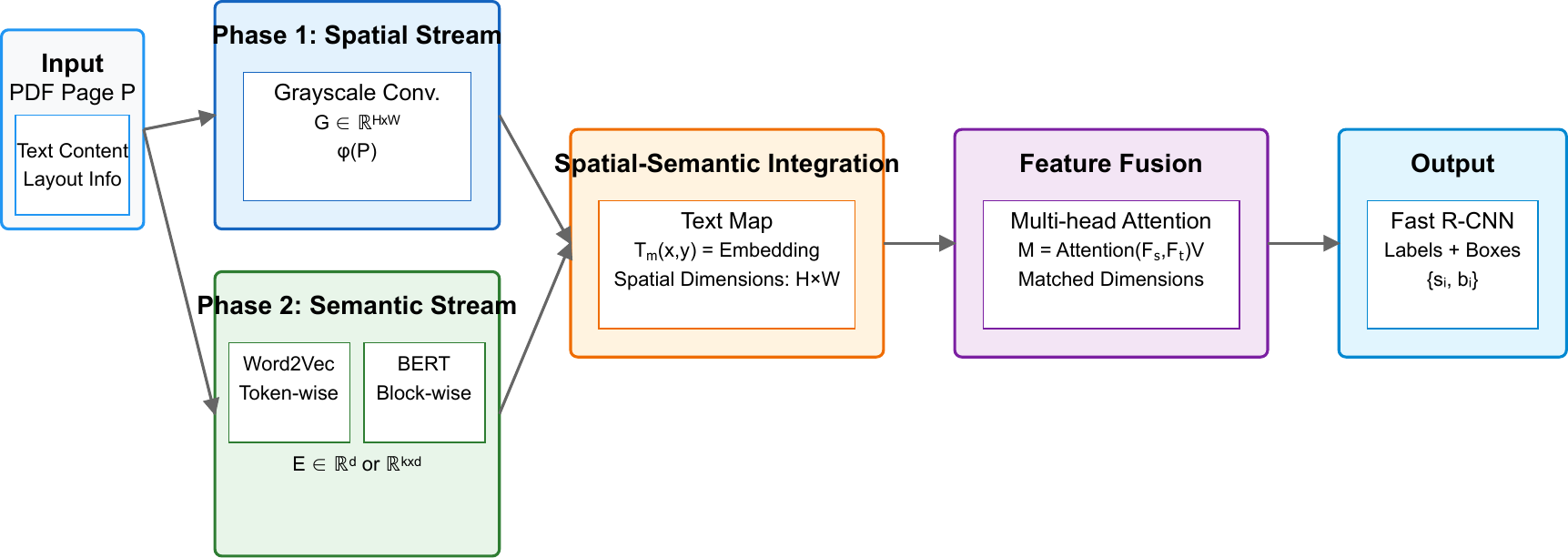}
    \caption{Overview of the TextMap approach}
    \label{fig:textmap}
\end{figure*}

\paragraph{Phase 1: Spatial Representation} transforms $\mathcal{P}$ into a grayscale representation $G$ that preserves structural information:
\begin{equation}
    G = \phi(\mathcal{P}) \in \mathbb{R}^{H \times W}
\end{equation}
where $H$ and $W$ are the height and width of the page, respectively. This transformation preserves the spatial distribution of text and structural elements across the document.

\paragraph{Phase 2: Semantic Mapping}
The semantic mapping differs based on the chosen embedding function $\psi$. For Word2Vec, each token $\omega_i$ is embedded individually:
\begin{equation}
E_i = \psi_{Word2Vec}(\omega_i) \in \mathbb{R}^d
\end{equation}
For BERT, entire text blocks $B_j = {\omega_1, ..., \omega_k}j$ are embedded together to capture contextual relationships:
\begin{equation}
E_j = \psi{BERT}(B_j) \in \mathbb{R}^{k \times d}
\end{equation}
where $d$ is the embedding dimension, and $k$ is the number of tokens in the block.

\subsubsection{Spatial-Semantic Integration}
The key innovation in our approach is the integration of spatial and semantic information through a carefully designed interpolation process:

1. \textbf{Region Identification}:
The regions of interest $R = {R_1, ..., R_k}$ are determined by the locations of text content in the document. Each region $R_i$ corresponds to a bounding box containing embedded text:
\begin{multline}
R_i = \{(x, y, w, h) \mid \\
\text{text content exists at } (x,y) \text{ with width } w \text{ and height } h\}
\end{multline}

The regions are naturally defined by the presence of text content that has been extracted and embedded. This ensures that our regions directly correspond to actual textual content in the document.

\paragraph{Embedding Interpolation}
For each region $R_i$, the embedding is directly mapped into the spatial coordinates of that region to create the text map $T_m$. We perform a straightforward mapping to ensure that each spatial location in the text map contains the embedding of the text (token or block) that appears at that location in the original document. For Word2Vec, where each token has its own embedding:
\begin{equation}
T_m(x,y) = E_j \quad \text{where} \quad (x,y) \in R_i \text{ contains token } \omega_j
\end{equation}
For BERT, where entire blocks are embedded together:
\begin{equation}
T_m(x,y) = E_i \quad \text{where} \quad (x,y) \in R_i \text{ contains block } B_i
\end{equation}

3. \textbf{Feature Fusion}:
After interpolation, both the grayscale representation $G$ and the text map $T_m$ have matching spatial dimensions, as the embeddings have been mapped to their corresponding regions' coordinates in the document space. Specifically:

$G \in \mathbb{R}^{H \times W}$ from the spatial stream
$T_m \in \mathbb{R}^{H \times W \times d}$ from the interpolated embeddings, where $d$ is the embedding dimension

This dimensional alignment allows us to apply convolutional operations to both streams:
\begin{equation}
F_{spatial} = Conv2D(G) \in \mathbb{R}^{H' \times W' \times C}
\end{equation}
\begin{equation}
F_{semantic} = Conv2D(T_m) \in \mathbb{R}^{H' \times W' \times C}
\end{equation}
where $H'$ and $W'$ are the reduced spatial dimensions after convolution, and $C$ is the number of output channels.
These spatially-aligned feature maps are then fused using a multi-head attention mechanism:
\begin{equation}
M = Attention(F_{spatial}, F_{semantic})V
\end{equation}
where $V$ is a learnable value matrix. This fusion process effectively combines the structural information from the spatial stream with the semantic information from the text embeddings, while maintaining spatial correspondence between the two streams.

\subsubsection{Segmentation and Classification}
The fused features $M$ are processed through a Fast R-CNN architecture for final segmentation and classification. For each identified region, we predict both the class label and bounding box refinements:
\begin{equation}
    \{s_i, b_i\} = FastRCNN(R_{refined})
\end{equation}
where $s_i$ is the metadata label and $b_i$ are the refined coordinates.

\subsubsection{Joint Optimization Framework}
The model is trained through a joint optimization framework that combines three objectives:

1. \textbf{Semantic Objective} ($\mathcal{L}_{semantic}$):
\begin{equation}
    \mathcal{L}_{semantic}(\theta) = -\sum_{i=1}^n \log P(s_i|\omega_i, E_i)
\end{equation}
This term ensures accurate metadata label assignment based on textual content.

2. \textbf{Spatial Objective} ($\mathcal{L}_{spatial}$):
\begin{equation}
    \mathcal{L}_{spatial}(\theta) = \sum_{i=1}^n \sum_{j \in \mathcal{N}(i)} \|f(G_i) - f(G_j)\|_2^2 \cdot \mathbb{1}[s_i = s_j]
\end{equation}
where $\mathcal{N}(i)$ represents the spatial neighbors of token $i$, $f$ is a feature extraction function, and $\mathbb{1}$ is the indicator function.

3. \textbf{Cross-modal Objective} ($\mathcal{L}_{cross}$):
\begin{equation}
    \mathcal{L}_{cross}(\theta) = -\sum_{i=1}^n \log P(s_i|E_i, f(G_i))
\end{equation}

The complete optimization objective is:
\begin{equation}
    \mathcal{J}(\theta) = \alpha\mathcal{L}_{semantic}(\theta) + \beta\mathcal{L}_{spatial}(\theta) + \gamma\mathcal{L}_{cross}(\theta)
\end{equation}
where $\alpha$, $\beta$, and $\gamma$ are learnable parameters that balance the contribution of each term.

\subsubsection{Training and Inference}
During training, we optimize $\mathcal{J}(\theta)$ using mini-batch stochastic gradient descent:
\begin{equation}
    \theta_{t+1} = \theta_t - \eta \nabla_\theta \mathcal{J}(\theta_t)
\end{equation}
where $\eta$ is the learning rate.

At inference time, for a new document page $\mathcal{P}$, we:
1. Generate spatial features $G = \phi(\mathcal{P})$
2. Compute embeddings $E_i = \psi(\omega_i)$ for each token
3. Identify regions
4. Apply the trained model to obtain metadata labels:
\begin{equation}
    s_i = \argmax_{l \in L} P(l|E_i, f(G_i))
\end{equation}

This formulation ensures that:
\begin{itemize}
    \item Tokens with similar semantics and spatial proximity are likely to share labels
    \item The model can handle variable document layouts
    \item Both local and global document structure are considered
    \item The extraction is robust to variations in formatting
\end{itemize}
\section{Experiments}
\label{sec:exp}
In this section, we compare the performance of the described methods in the previous section on two challenging datasets.

\subsection{Dataset}
This section presents a comparative analysis of the different methodologies aimed at extracting metadata from academic PDF documents. To this end, we prepared two challenging datasets, namely SSOAR-MVD and S-PMRD.

\subsubsection{SSOAR Multidisciplinary Vision Dataset (SSOAR-MVD)} 

To ensure a fair comparison of all the methods described in the section, we ensured they were all applied to the same dataset. As a result, we collected a challenging dataset of 50,000 documents from the SSOAR repositor\footnote{\url{https://www.gesis.org/en/ssoar/home}}. The SSOAR stores publications from various publishers, including small and mid-sized ones, covering a range of disciplines known for their challenging layout formats, such as Social Sciences, Humanities, Law, and Administration. This guarantees that a wide variety of templates are included in the dataset. During the scraping process, each document was downloaded along with its textual metadata provided by the SSOAR repository. However, since most computer vision approaches require labelled images (i.e., bounding boxes), a preprocessing phase was conducted to ensure this. Each document underwent the following steps:

\begin{itemize}
    \item The document is converted into an image.
    \item Using an open-source tool provided by TensorFlow, blocks of text were extracted from the document along with their respective bounding boxes.
    \item The similarity between each text block and metadata class was measured (using the collected textual metadata from the SSOAR).
    \item If a certain block had a near-perfect similarity with a specific class, the corresponding bounding box was assigned to that class. Otherwise, it was assigned a class "other".
\end{itemize}

\subsubsection{S2ORC PDF Metadata Refinement Dataset (S-PMRD)} 

To evaluate the efficacy of these methods on an authentic corpus, we meticulously curated a subset from the Semantic Scholar Open Research Corpus (S2ORC)~\cite{lo2019s2orc}. While S2ORC offers a vast repository of millions of scholarly articles, our preliminary analyses revealed significant discrepancies between the raw textual data provided by S2ORC and the content within the corresponding PDF documents. Notably, certain text segments available in the S2ORC dataset were absent from the PDFs, and this does not explain the extraction methodologies employed by S2ORC. These inconsistencies pose substantial challenges for conducting detailed academic analyses that necessitate precise text alignment, such as citation context analysis, text-based data mining, and metadata extraction.

To address these challenges, we developed a specialized sub-corpus of S2ORC, specifically aimed at extracting metadata directly from PDF documents, thereby bypassing the potential inaccuracies inherent in pre-extracted text. The processing pipeline for each document included in this sub-corpus is as follows:

\begin{itemize}
    \item Using the Digital Object Identifier (DOI) from S2ORC, additional metadata, including links to the actual PDF documents, is retrieved from CrossRef~\cite{hendricks2020crossref} using their API.
    \item PDFs are downloaded when available, acknowledging that some links may be inactive or access-restricted.
    \item Text is extracted from the first page of each PDF. This extracted text undergoes a normalization process to eliminate irregularities such as inconsistent spacing and line breaks, thereby ensuring uniformity across the dataset.
    \item We employ both exact and fuzzy matching techniques to extract critical metadata elements, including author names, titles, abstracts, and affiliations. This dual-method approach accommodates minor discrepancies due to text recognition errors or formatting variations.
    \item Each document is converted into an image representation.
    \item For each identified metadata element, bounding boxes are determined based on their locations within the text.
\end{itemize}

Ultimately, each instance in this dataset comprises:

\begin{itemize}
    \item The original PDF file.
    \item The normalized text extracted from the PDF.
    \item An image of the first page of the PDF.
    \item Metadata attributes annotated with both their positions in the extracted text and their coordinates on the corresponding image.
\end{itemize}

\section{Settings}

In this study, we employ a token-level evaluation to validate each model's effectiveness, where each predicted token is compared against the ground truth annotations in our dataset. This evaluation is conducted using standard metrics such as Precision, Recall, and F1-Score. 

\subsection{Results}

In the first analysis, we evaluate the performance of the presented methods on the datasets SSOAR-MVD and S-PMRD, as depicted in Tables Tables~\ref{tab:ev1:ssoar} and \ref{tab:ev1:pmrd}, illustrating the outcomes in terms of Precision, Recall, and F1-Score to provide a multi-dimensional comparison. Notably, the proposed TextMap-Word2Vec method achieves the highest F1-score of 0.913, suggesting that capturing both the semantics of the PDF content and its layout is particularly effective for this task. This is evident by the performance of the Fast-RCNN and Vision-Langauge methods which also leverage the layout and the appearance of the PDF document. In contrast, traditional models like CRF exhibit lower performance metrics, which may indicate difficulties in adapting to the multi-faceted nature of PDF content, where layout and semantic context play crucial roles. An important observation is that the embedding approach in TextMap plays a significant role in the performance of the model. 

\begin{table*}[ht!]
\centering
\caption{Overall Performance Comparison of Different Methods on SSOAR-MVD}
\label{tab:ev1:ssoar}
\begin{tabular}{|l|c|c|c|c|c|}
\hline
\textbf{Method} & \textbf{Precision Macro} & \textbf{Precision Micro} & \textbf{Recall Macro} & \textbf{Recall Micro} & \textbf{F1-score} \\ \hline
CRF & 0.609 & 0.544 & 0.524 & 0.471 & 0.57 \\ \hline
BiLSTM & 0.901 & 0.89 & 0.898 & 0.861 & 0.9 \\ \hline
LSTM-CRF & 0.778 & 0.713 & 0.745 & 0.697 & 0.761 \\ \hline
GROBID & 0.854 & 0.671 & 0.794 & 0.551 & 0.821 \\ \hline
Fast-RCNN & 0.9 & 0.915 & 0.896 & 0.904 & 0.898 \\ \hline
%the results of the model below need to be verified 
Vision-Language & 0.935 & \textbf{0.94} & 0.902 & 0.904 & 0.92 \\ \hline
TextMap-Bert & 0.908 & 0.887 & 0.902 & 0.897 & 0.905 \\ \hline
TextMap-Word2Vec & \textbf{0.917} & 0.92 & \textbf{0.91} & \textbf{0.904} & \textbf{0.913} \\ \hline
TextMap-Char2Vec & 0.845 & 0.8 & 0.849 & 0.797 & 0.847 \\ \hline
\end{tabular}
\end{table*}

\begin{table*}[ht!]
\centering
\caption{Overall Performance Comparison of Different Methods on S-PMRD}
\label{tab:ev1:pmrd}
\begin{tabular}{|l|c|c|c|c|c|}
\hline
\textbf{Method} & \textbf{Precision Macro} & \textbf{Precision Micro} & \textbf{Recall Macro} & \textbf{Recall Micro} & \textbf{F1-score} \\ \hline
CRF & 0.573 & 0.521 & 0.501 & 0.45 & 0.511 \\ \hline
BiLSTM & 0.883 & 0.872 & 0.898 & 0.863 & 0.889 \\ \hline
LSTM-CRF & 0.740 & 0.707 & 0.736 & 0.692 & 0.724 \\ \hline
GROBID & 0.822 & 0.651 & 0.787 & 0.542 & 0.791 \\ \hline
Fast-RCNN & 0.874 & 0.906 & 0.886 & 0.912 & 0.893 \\ \hline
Vision-Language & 0.91 & \textbf{0.923} & 0.904 & 0.911 & 0.903 \\ \hline
TextMap-Bert & 0.882 & 0.860 & 0.916 & 0.887 & 0.894 \\ \hline
TextMap-Word2Vec & \textbf{0.892} & 0.902 & \textbf{0.91} & \textbf{0.902} & \textbf{0.901} \\ \hline
TextMap-Char2Vec & 0.815 & 0.786 & 0.841 & 0.792 & 0.821 \\ \hline
\end{tabular}
\end{table*}

To give a better overview of the performance of each method, we present below the result of each method for each attribute. Table~\ref{tab:ev2:CRF-SSOAR} presents the detailed results of CRF on SSOAR-MVD. As demonstrated, CRF excels in attributes with structured and predictable formats such as Dates (F1-score: $0.731$) and DOIs (F1-score: $0.749$), where the patterned nature of the data plays to the strengths of the CRF's sequence modelling capabilities. However, CRF struggles with more complex and less structured text such as Titles (F1-score: $0.433$) and Abstracts (F1-score: $0.392$), where the variability in content and formatting challenges its ability to accurately predict boundaries and content. The moderate success in extracting Authors (F1-score: $0.482$) and higher performance in Affiliation data (F1-score: $0.672$) suggest that CRF can handle semi-structured text effectively when patterns are somewhat predictable.

\begin{table}[ht!]
\centering
\caption{Performance Metrics for CRF Method on SSOAR-MVD}
\label{tab:ev2:CRF-SSOAR}
\begin{tabular}{|l|c|c|c|}
\hline
\textbf{Category} & \textbf{Precision Macro} & \textbf{Recall Macro} & \textbf{F1-score} \\ \hline
Title & 0.568 & 0.35 & 0.433 \\ \hline
Abstract & 0.457 & 0.344 & 0.392 \\ \hline
Authors & 0.57 & 0.418 & 0.482 \\ \hline
Email & 0.612 & 0.607 & 0.609 \\ \hline
Address & 0.522 & 0.481 & 0.5 \\ \hline
Date & 0.754 & 0.71 & 0.731 \\ \hline
Journal & 0.547 & 0.577 & 0.561 \\ \hline
Affiliation & 0.663 & 0.682 & 0.672 \\ \hline
DOI & 0.795 & 0.709 & 0.749 \\ \hline
Macro Average & 0.609 & 0.524 & 0.57 \\ \hline
Micro Average & 0.544 & 0.471 & N/A \\ \hline
\end{tabular}
\end{table}

Table~\ref{tab:ev2:BILSTM-SSOAR} presents the detailed results of Bi-LSTM on SSOAR-MVD. The BiLSTM method exhibits strong performance across several metadata categories such as `Title', `Author', etc. This reflects its robust capability to capture both the context and sequence of text within scholarly PDF documents. Specifically, the method maintains strong performance in handling Abstracts and Dates, with F1-scores slightly above 0.910. This indicates BiLSTM's adeptness at managing narrative content and specific formatted text. However, the slightly lower performance in extracting Affiliation data, with an F1-score of $0.833$, suggests some challenges in dealing with categories of short strings.

\begin{table}[ht!]
\centering
\caption{Performance Metrics for BiLSTM Method on SSOAR-MVD}
\label{tab:ev2:BILSTM-SSOAR}
\begin{tabular}{|l|c|c|c|}
\hline
\textbf{Category} & \textbf{Precision Macro} & \textbf{Recall Macro} & \textbf{F1-score} \\ \hline
Title & 0.931 & 0.91 & 0.920 \\ \hline
Abstract & 0.908 & 0.914 & 0.911 \\ \hline
Authors & 0.944 & 0.93 & 0.937 \\ \hline
Email & 0.881 & 0.86 & 0.870 \\ \hline
Address & 0.9 & 0.882 & 0.891 \\ \hline
Date & 0.916 & 0.905 & 0.910 \\ \hline
Journal & 0.865 & 0.891 & 0.878 \\ \hline
Affiliation & 0.814 & 0.853 & 0.833 \\ \hline
DOI & 0.952 & 0.94 & 0.946 \\ \hline
Macro Average & 0.901 & 0.898 & 0.9 \\ \hline

Micro Average & 0.89 & 0.861 & N/A \\ \hline
\end{tabular}
\end{table}

Table~\ref{tab:ev2:LSTM-CRF:SSOAR} presents the detailed results of LSTM-CRF on SSOAR-MVD. The LSTM-CRF method demonstrates moderate efficacy in extracting different metadata categories, notably outperforming traditional CRF models that rely on handcrafted features. This enhancement suggests that the LSTM architecture provides a more robust feature representation for CRF to utilize effectively. However, despite its competencies across various categories, LSTM-CRF does not surpass the BiLSTM method, which exhibits superior handling of long-term sequential and contextual data dependencies.

\begin{table}[ht!]
\centering
\caption{Performance Metrics for LSTM-CRF Method on SSOAR-MVD}
\label{tab:ev2:LSTM-CRF:SSOAR}
\begin{tabular}{|l|c|c|c|}
\hline
\textbf{Category} & \textbf{Precision Macro} & \textbf{Recall Macro} & \textbf{F1-score} \\ \hline
Title & 0.741 & 0.699 & 0.719 \\ \hline
Abstract & 0.688 & 0.7 & 0.693 \\ \hline
Authors & 0.84 & 0.815 & 0.827 \\ \hline
Email & 0.801 & 0.782 & 0.791 \\ \hline
Address & 0.725 & 0.76 & 0.742 \\ \hline
Date & 0.89 & 0.822 & 0.854 \\ \hline
Journal & 0.739 & 0.727 & 0.732 \\ \hline
Affiliation & 0.774 & 0.68 & 0.723 \\ \hline
DOI & 0.81 & 0.724 & 0.764 \\ \hline
Macro Average & 0.778 & 0.745 & 0.761 \\ \hline

Micro Average & 0.713 & 0.697 & N/A \\ \hline
\end{tabular}
\end{table}

Table~\ref{tab:ev2:Grobid:SSOAR} presents the detailed results of Grobid on SSOAR-MVD. Despite its simple design, GROBID exhibits robust performance across different categories. It particularly excels in extracting Authors and Abstracts, achieving impressive F1 scores of 0.958 and 0.935 respectively. These high scores can be attributed to GROBID’s effective application of its cascading sequence labelling models, which are adept at handling well-structured and clearly delineated data. However, GROBID encounters some variability in categories involving more complex or less standardized information, such as Address and Affiliation. This variability stems from these categories' inherent challenges, including inconsistent formatting and multifaceted data structures that can complicate the parsing process. While the cascading approach of GROBID generally enhances its capability to manage hierarchical document structures efficiently, it occasionally struggles with elements that lack a clear or uniform presentation.

\begin{table}[ht!]
\centering
\caption{Performance Metrics for GROBID Method on SSOAR-MVD}
\label{tab:ev2:Grobid:SSOAR}
\begin{tabular}{|l|c|c|c|}
\hline
\textbf{Category} & \textbf{Precision Macro} & \textbf{Recall Macro} & \textbf{F1-score} \\ \hline
Title & 0.764 & 0.667 & 0.951 \\ \hline
Abstract & 0.84 & 0.79 & 0.935 \\ \hline
Authors & 0.934 & 0.855 & 0.958 \\ \hline
Email & 0.91 & 0.812 & 0.893 \\ \hline
Address & 0.722 & 0.78 & 0.872 \\ \hline
Date & 0.855 & 0.877 & 0.873 \\ \hline
Journal & 0.887 & 0.75 & 0.927 \\ \hline
Affiliation & 0.859 & 0.813 & 0.818 \\ \hline
DOI & 0.911 & 0.8 & 0.916 \\ \hline
Macro Average & 0.854 & 0.794 & 0.821 \\ \hline

Micro Average & 0.671 & 0.551 & N/A \\ \hline
\end{tabular}
\end{table}

Table~\ref{tab:ev2:Fast-RCNN:SSOAR} presents the detailed results of Fast-RCNN on SSOAR-MVD. The model demonstrates high precision and recall across most categories, with solid performance in the Title, Abstract, and Journal categories, suggesting robustness in recognizing well-defined metadata fields. The Email and Authors categories also show commendable accuracy. However, the model indicates slightly weaker performance in the Address and Date categories, with F1-scores of 0.837 and 0.832, respectively. This could be due to variability in the formatting and presentation of these metadata elements across documents, which poses challenges in consistent extraction. The small variance between Macro Average and Micro Average metrics indicates a balanced performance across different categories, without significant bias toward any particular type of metadata.

\begin{table}[ht!]
\centering
\caption{Performance Metrics for Fast-RCNN Method on SSOAR-MVD}
\label{tab:ev2:Fast-RCNN:SSOAR}
\begin{tabular}{|l|c|c|c|}
\hline
\textbf{Category} & \textbf{Precision Macro} & \textbf{Recall Macro} & \textbf{F1-score} \\ \hline
Title & 0.966 & 0.95 & 0.958 \\ \hline
Abstract & 0.915 & 0.922 & 0.918 \\ \hline
Authors & 0.91 & 0.938 & 0.924 \\ \hline
Email & 0.933 & 0.917 & 0.925 \\ \hline
Address & 0.875 & 0.802 & 0.837 \\ \hline
Date & 0.825 & 0.84 & 0.832 \\ \hline
Journal & 0.94 & 0.925 & 0.932 \\ \hline
Affiliation & 0.839 & 0.876 & 0.857 \\ \hline
DOI & 0.901 & 0.893 & 0.897 \\ \hline
Macro Average & 0.9 & 0.896 & 0.898 \\ \hline

Micro Average & 0.915 & 0.904 & N/A \\ \hline
\end{tabular}
\end{table}

Tables~\ref{tab:ev2:TextMap-Bert:SSOAR}, \ref{tab:ev2:TextMap-Word2Vec:SSOAR}, \ref{tab:ev2:TextMap-Char2Vec:SSOAR} present the performance metrics across three different configurations of the TextMap model, using BERT, Word2Vec, and Char2Vec embeddings. we can observe varied performances that highlight the strengths and weaknesses of each embedding strategy with the TextMap approach. TextMap using BERT Embeddings configuration (Table~\ref{tab:ev2:TextMap-Bert:SSOAR}) demonstrates strong performance across several categories, particularly in Authors, Abstract, and Journal, with F1-scores $>0.92$. This suggests that BERT's deep contextual embeddings are particularly effective at extracting structured text like titles and authors' details, typically well-defined in the document. The lower performance in the Affiliation and Address categories, with relatively lower F1 scores, could indicate challenges in capturing less consistently formatted information.

TextMap using Word2Vec Embeddings configuration (Table~\ref{tab:ev2:TextMap-Word2Vec:SSOAR}) generally performs well, particularly in the Authors and Abstract categories. This suggests good generalization in capturing both the semantic and structural patterns in data, though slightly less effectively than BERT in terms of overall averages. However, as shown in Table~\ref{tab:ev:comp}, it has a lower computational cost in both training and inference. Consequently, this configuration provides a good balance between performance and computational efficiency, especially suitable for environments where computational resources or training data are limited.

TextMap using Char2Vec Embeddings configuration (Table~\ref{tab:ev2:TextMap-Char2Vec:SSOAR}) demonstrates a notable decline in performance across most categories compared to the BERT and Word2Vec models. It performs best in the DOI category with an F1-score $>0.9$ but struggles particularly with Abstract and Journal metadata, with F1-scores $<0.8$. In conclusion, Char2Vec, while useful in certain niche applications (like OCR and typo-sensitive extractions), may not be suitable for tasks requiring deep semantic understanding such as understanding the semantics of a scholarly text.

\begin{table}[ht!]
\centering
\caption{Performance Metrics for TextMap using Bert embeddings.}
\label{tab:ev2:TextMap-Bert:SSOAR}
\begin{tabular}{|l|c|c|c|}
\hline
\textbf{Category} & \textbf{Precision Macro} & \textbf{Recall Macro} & \textbf{F1-score} \\ \hline
Title & 0.954 & 0.949 & 0.951 \\ \hline
Abstract & 0.921 & 0.95 & 0.935 \\ \hline
Authors & 0.967 & 0.949 & 0.958 \\ \hline
Email & 0.91 & 0.877 & 0.893 \\ \hline
Address & 0.889 & 0.856 & 0.872 \\ \hline
Date & 0.855 & 0.891 & 0.873 \\ \hline
Journal & 0.924 & 0.93 & 0.927 \\ \hline
Affiliation & 0.822 & 0.815 & 0.818 \\ \hline
DOI & 0.931 & 0.902 & 0.916 \\ \hline
Macro Average & 0.908 & 0.902 & 0.905 \\ \hline

Micro Average & 0.887 & 0.897 & N/A \\ \hline
\end{tabular}
\end{table}

\begin{table}[ht!]
\centering
\caption{Performance Metrics for TextMap using Word2Vec embeddings}
\label{tab:ev2:TextMap-Word2Vec:SSOAR}
\begin{tabular}{|l|c|c|c|}
\hline
\textbf{Category} & \textbf{Precision Macro} & \textbf{Recall Macro} & \textbf{F1-score} \\ \hline
Title & 0.962 & 0.922 & 0.941 \\ \hline
Abstract & 0.933 & 0.952 & 0.942 \\ \hline
Authors & 0.978 & 0.949 & 0.963 \\ \hline
Email & 0.93 & 0.899 & 0.914 \\ \hline
Address & 0.904 & 0.86 & 0.881 \\ \hline
Date & 0.852 & 0.907 & 0.878 \\ \hline
Journal & 0.924 & 0.94 & 0.931 \\ \hline
Affiliation & 0.834 & 0.849 & 0.841 \\ \hline
DOI & 0.933 & 0.915 & 0.923 \\ \hline
Macro Avrerage & 0.917 & 0.91 & 0.913 \\ \hline

Micro Average & 0.92 & 0.904 & N/A \\ \hline
\end{tabular}
\end{table}

\begin{table}[ht!]
\centering
\caption{Performance Metrics for TextMap using Char2Vec embeddings}
\label{tab:ev2:TextMap-Char2Vec:SSOAR}
\begin{tabular}{|l|c|c|c|}
\hline
\textbf{Category} & \textbf{Precision Macro} & \textbf{Recall Macro} & \textbf{F1-score} \\ \hline
Title & 0.851 & 0.874 & 0.862 \\ \hline
Abstract & 0.77 & 0.8 & 0.785 \\ \hline
Authors & 0.849 & 0.815 & 0.832 \\ \hline
Email & 0.902 & 0.89 & 0.896 \\ \hline
Address & 0.86 & 0.881 & 0.870 \\ \hline
Date & 0.875 & 0.842 & 0.858 \\ \hline
Journal & 0.782 & 0.809 & 0.795 \\ \hline
Affiliation & 0.804 & 0.83 & 0.817 \\ \hline
DOI & 0.917 & 0.905 & 0.911 \\ \hline
Macro Average & 0.845 & 0.849 & 0.847 \\ \hline

Micro Average & 0.8 & 0.797 & N/A \\ \hline
\end{tabular}
\end{table}

In addition to comparing the models' performance in terms of precision, recall, and F1 score, we compare their computational complexity using the SSOAR-MVD dataset.

This comparison reveals a range of trade-offs between computational efficiency and performance accuracy. CRF offers the quickest inference time and requires the least training time, making them ideal for environments where speed is prioritized over cutting-edge accuracy. BiLSTM models, while requiring more extensive training, provide rapid inference capabilities, suitable for real-time applications once the model is deployed. LSTM-CRF models combine the deep learning prowess of LSTMs with the structured output of CRFs did not achieve higher accuracy compared to Bi-LSTM models and has longer training times and moderately slow inference speeds. GROBID, tailored specifically for document processing tasks, demands the most extended training period and exhibits slower inference times, reflecting its comprehensive analytical depth. Fast R-CNN, effective in precise localization of content within documents, also shows a moderate training duration with slower inference, suited to applications where precision is more critical than speed. Vision-language models, though offering superior performance where an understanding of both visual cues and textual information is necessary, involve the longest training durations and the slowest inference rates, which could be a significant drawback in time-sensitive scenarios. 

Lastly, TextMap models using BERT, Word2Vec, and Char2Vec embeddings demonstrate a spectrum of efficiencies, with BERT providing high accuracy but slower inference and longer training times, whereas Word2Vec offers a more balanced approach, making it preferable for scenarios that demand both efficiency and effectiveness.

\begin{table}[]

    \centering
    \caption{Average Training and Inference Times for Different Machine Learning Models Used in Metadata Extraction. The table lists the estimated average training and inference times for each model and standard deviations for these estimates.}
    \label{tab:ev:comp}
    \begin{tabular}{c|c|c}
        Model & Training time & Inference time\\
        CRF  &  $36 \pm 7.2$ hours & $0.5 \pm 0.01$ seconds \\
        BiLSTM & $84 \pm 7.1$ hours & $0.1$ seconds\\
        LSTM-CRF & $126 \pm 3.7$ hours & $0.2$ seconds\\
        GROBID & $156 \pm 7.2$ hours & $1.2 \pm 0.6$ seconds\\
        Fast-RCNN & $60 \pm 6.8$ hours & $1.3 \pm 0.4$ seconds\\
        Vision-Language &  $192 \pm 14.5$ hours & $3.5 \pm 1.11$ seconds\\
        TextMap-Bert &  $172 \pm 13.3$ hours &  $1.3 \pm 0.58$\\
        TextMap-Word2Vec & $92 \pm 6.2$ & $0.4$ seconds \\
        TextMap-Char2Vec & $90 \pm 7.0$ & $0.4$ seconds \\
        
    \end{tabular}
    
\end{table}

\subsection{Limitations}

While the models examined in this study demonstrate considerable potential for metadata extraction from scholarly PDF documents, several limitations must be acknowledged to fully appreciate their applicability and scope of use.

\begin{itemize}
    \item \textbf{Dependency on Training Data}: All models, particularly deep learning-based ones like BiLSTM, LSTM-CRF, and TextMap with BERT embeddings, exhibit a high dependency on the quantity and quality of the training data. Their performance is contingent upon the availability of large, annotated datasets. This reliance can limit their practical deployment in scenarios where such datasets are not readily available or are too domain-specific.

    \item \textbf{Adaptability to Rapid Changes}: The field of digital publishing is evolving, with new standards and formats emerging. The adaptability of these models to such rapid changes has not been thoroughly tested, raising concerns about their long-term viability without continuous updates and retraining.

    \textbf{Error Propagation}: In multi-stage models like GROBID or Fast-RCNN, errors in early processing stages can propagate, leading to compounded errors in metadata extraction outcomes. This cascade effect can significantly affect the overall quality of the extracted metadata.

    \textbf{}
\end{itemize}
\section{Conclusion}
\label{sec:con}

This study has conducted a comprehensive comparison of various machine learning models to evaluate their effectiveness in extracting metadata from two challenging datasets. The analysis revealed significant variations in performance and computational demands across the models, underscoring the importance of selecting an appropriate model and architecture tailored to specific use case requirements.

The CRF and BiLSTM models demonstrated rapid inference capabilities coupled with robust performance, making them ideal candidates for real-time applications. In contrast, the LSTM-CRF hybrid model, despite combining the strengths of LSTMs and the structured output capabilities of CRFs, did not achieve results on par with its component technologies.

Models that integrate vision and language modalities, while resource-intensive, deliver depth and precision in analysis that simpler models cannot achieve. This sophistication makes them particularly valuable in scenarios where the accuracy of extracted metadata critically impacts the outcomes of subsequent processes.

The TextMap models, which leverage various embeddings such as BERT, Word2Vec, and Char2Vec, offer a spectrum of choices balancing training and inference times with performance. Among these, BERT embeddings stand out for their exceptional accuracy, albeit at a higher computational cost, illustrating the fundamental trade-offs between resource investment and extraction efficacy.

Ultimately, the selection of a metadata extraction model should be driven not only by dataset characteristics but also by the practical constraints of the use case—available computational resources, required inference speed, and the trade-offs between precision and performance that stakeholders are prepared to accept. Future research should consider the potential of hybrid models and the development of more efficient training algorithms to further optimize the application of machine learning in metadata extraction tasks, enhancing both their efficiency and accessibility.

\bibliographystyle{IEEEtran}
\bibliography{ref}
\iffalse
\begin{IEEEbiography}
Authors' biographies. 

[{\includegraphics[width=1in,height=1.25in,clip,keepaspectratio]{Figures/zeyd-modified.png}}]{Zeyd Boukhers}
leads the FAIR Data and Distributed Analytics group at the Fraunhofer Institute for Applied Information Technology (FIT). His research focuses on machine learning, computer vision, data science, natural language processing, and knowledge management. He earned his Ph.D. in Pattern Recognition from the University of Siegen, Germany.
\end{IEEEbiography}

\begin{IEEEbiography}[{\includegraphics[width=1in,height=1.25in,clip,keepaspectratio]{Figures/cong-modified.jpg}}]{Cong Yang}
is an Associate Professor in Soochow
University since 2022. Before that, he was a Postdoc
researcher at the MAGRIT team in INRIA (France).
Later, he worked scientifically and led the computer
vision and machine learning teams in Clobotics
and Horizon Robotics. His main research interests
are computer vision, pattern recognition and their
interdisciplinary applications. Cong earned his Ph.D.
degree in computer vision and pattern recognition
from the University of Siegen (Germany) in 2016.

\end{IEEEbiography}
\fi

\end{document}